\documentclass[aps,prb,reprint,superscriptaddress,showpacs, showkeys, floatfix]{revtex4-1}

\usepackage{graphicx}
\usepackage{bm}
\usepackage{color}

\draft 

\begin{document}





\title{Transport Gap in Suspended Bilayer Graphene at Zero Magnetic Field}








\author{A.~Veligura}
\email[a.veligura@rug.nl]{}
\affiliation{Physics of Nanodevices,
Zernike Institute for Advanced Materials, University of Groningen,
Nijenborgh 4, 9747 AG Groningen,The Netherlands}

\author{H.J.~van~Elferen}
\affiliation{High Field Magnet Laboratory and Institute for
Molecules and Materials, Radboud University Nijmegen, Toernooiveld
7, 6525 ED Nijmegen, The Netherlands}

\author{N.~Tombros}
\affiliation{Physics of Nanodevices, Zernike Institute for Advanced
Materials, University of Groningen, Nijenborgh 4, 9747 AG
Groningen,The Netherlands}

\author{J.C.~Maan}
\affiliation{High Field Magnet Laboratory and Institute for
Molecules and Materials, Radboud University Nijmegen, Toernooiveld
7, 6525 ED Nijmegen, The Netherlands}

\author{U.~Zeitler}
\affiliation{High Field Magnet Laboratory and Institute for
Molecules and Materials, Radboud University Nijmegen, Toernooiveld
7, 6525 ED Nijmegen, The Netherlands}

\author{B.J.~van~Wees}
\affiliation{Physics of Nanodevices, Zernike Institute for Advanced
Materials, University of Groningen, Nijenborgh 4, 9747 AG
Groningen,The Netherlands}






\date{\today}

\begin{abstract}

We report a change of three orders of magnitudes in the resistance
of a suspended bilayer graphene flake which varies from a few
k$\Omega$s in the high carrier density regime to several M$\Omega$s
around the charge neutrality point (CNP). The corresponding
transport gap is 8~meV at 0.3~K. The sequence of appearing quantum
Hall plateaus at filling factor $\nu=2$ followed by $\nu=1$ suggests
that the observed gap is caused by the symmetry breaking of the
lowest Landau level. Investigation of the gap in a tilted magnetic
field indicates that the resistance at the CNP
shows a weak linear decrease for increasing total magnetic field. Those observations are in agreement with a spontaneous valley
splitting at zero magnetic field followed by splitting of the spins
originating from different valleys with increasing magnetic field.
Both, the transport gap and $B$ field response point
toward spin polarized layer antiferromagnetic state as a ground
state in the bilayer graphene sample. The observed non-trivial
dependence of the gap value on the normal component of $B$ suggests
possible exchange mechanisms in the system.

\end{abstract}


\maketitle 



\section{Introduction}
Followed by the isolation of single layer graphene, the study of
bilayer graphene (BLG) became a separate direction of research in
the community of two dimensional materials. Charge carriers in
bilayer graphene have a parabolic dispersion with an effective mass
of about 0.054m$_e$,\cite{Mccannbilayers, McCann} but also possess a
chirality. The latter manifests itself in an unconventional quantum
Hall effect\cite{Novoselov_UQHE} with the lowest Landau level (LLL)
being eight fold degenerate. Compared to single layer, bilayer
graphene has next, to spin and valleys degrees of freedom, an
additional orbital degree of freedom, where Landau levels with
numbers n~=~0 and 1 (each four fold degenerate) have the same
energy.\cite{Novoselov_UQHE,McCann} Recent advances in obtaining
suspended bilayer graphene devices with charge carrier mobility
exceeding $\mu > 10,000~\text{cm}^2\text{V}^{-1}\text{s}^{-1}$ gave
access to the investigation of many-body phenomena in clean bilayer
graphene at low charge carrier concentration ($n <
10^{10}~\text{cm}^{-2}$).\cite{Feldman1,Weitz1,Martin1,Velasco1,Elferen,Freitag,
Mayorov,Bao3}

Due to the non vanishing density of states at the charge neutrality
point (CNP), bilayer graphene is predicted to have a variety of
ground states triggered by electron-electron interaction. There are
two competing theories describing the ground state of BLG: a
transition (i) to a gapped layer polarized state (excitonic
instability)\cite{Nandkishore1,Nandkishore2,Min,Zhang1,Jung1,Zhang2}
or (ii) to a gapless nematic phase.\cite{Vafek2, Lemonik, Toke}

Excitonic instability is a layer polarization in which the charge
density contribution from each valley and spin spontaneously shifts
to one of the two graphene layers.\cite{Jung1,Zhang2} This
redistribution is caused by an arbitrarily weak interaction between
charge from conduction and valence band states.\cite{Nandkishore1,
Nandkishore2} Since each bilayer flavor (spin or valley) can
polarize towards either of the two layers, there are 16 possible
states,\cite{Jung1,Zhang2} which can be classified by the total
polarization as being $layer$ ferromagnetic (all degrees of freedom
choose the same layer), $layer$ ferrimagnetic (three of the four
valley-spin flavors choose the same layer), or $layer$
antiferromagnetic (with no overall polarization). To make it clear,
the therm $"magnetic"$ should be associated to flavors (not only
spin) orientation in between two layers. These states are considered
as analogous to the biased bilayer\cite{Castro} in the sense that
the charge transfer can be attributed to the (wave vector dependent)
exchange potential difference between low-energy sites on the
opposite layers.\cite{Jung1} The total energy of the system is
lowered by the gain in the exchange interaction via breaking of the
inversion symmetry, i.e. introducing a gapped state.
Antiferromagnetic polarization is electrostatically favorable due to
the absence of a net charge on both layers, however, the actual
ground state is theoretically
undefined.\cite{Nandkishore1,Zhang2,Jung1} Recent experiments have
suggested the evidence of the possible existence of two of the
antiferromagnetic states - the anomalous quantum Hall state
(AQH)\cite{Weitz1,Martin1} and spin polarized layer
antiferromagnetic state (LAF)\cite{Velasco1}. To avoid
possible confusion we note that in earlier literature\cite{Jung1}
the LAF state is also called quantum valley Hall state. The AQH has
electrons being polarized in the same layer for both spins and in
opposite layers for opposite valleys.\cite{Zhang2,Jung1} This state
has spontaneously broken time reversal symmetry and therefore
possess a substantial orbital magnetization exhibiting quantized
Hall effect (at zero magnetic field), while its spin density is
everywhere zero.\cite{Zhang2} Due to its magnetization the AQH can
be favored over other ground states in the perpendicular magnetic
field. The LAF state has opposite spin-polarization for opposite
layers. In contrast to AQH, the LAF state does not have
topologically protected edge states, which brings its minimum
conductance to zero. For both states the theoretical estimations of
the gap $\Delta$ give the value of 1.5-30~meV.\cite{Nandkishore2,
Jung1} However, the inter-valley exchange weakly favors
the LAF state.\cite{Jung1,Zhang3} One of the ways to determine the
character of the bilayer ground state experimentally is to
investigate the response of the gap value to the magnetic field $B$
(which couples to spin) and electrical field $E$ (which couples to
layer pseudospin).\cite{Zhang3} When Zeeman coupling is included,
the QAH state quasiparticles simply spin-split, leaving the ground
state unchanged but the charge gap reduced. It was calculated that
for a 4~meV spontaneous gap at zero-field, a field of $B$~=~35~T
drives the gap to zero. On the other hand, the gap of LAF is
weakly $B$ field dependent.

The second possible description for the ground state of BLG is based
on a nematic phase caused by the renormalization of the low energy
spectrum.\cite{Vafek2, Lemonik} Detailed tight-binding model studies
showed that including next-neighbor interlayer coupling changes the
band structure in bilayer producing a Lifshitz transition in which
the isoenergetic line about each valley is broken into four pockets
with linear dispersion.\cite{McCann, McCann2} At the energies higher
then 1~meV the four pockets merge into one pocket with the usual
quadratic dispersion. Moreover, electron-electron interactions might
result in the further energy spectrum transformation, where the
number of low energy cones can be reduced to two near each of the
two $K$ points.\cite{Vafek2, Lemonik} In this case the minimum
conductance of the bilayer graphene is supposed to be increased
comparing to bilayer with parabolic dispersion ($8e^2/h$). This
scenario was also supported by the experimental result on the
suspended bilayer graphene in which strong spectrum reconstructions
and electron topological transitions were observed.\cite{Mayorov}

 In this paper we present electric transport properties of suspended
bilayer graphene by studying its behavior in tilted magnetic fields.
At $B$~=~0~T we observe the spontaneous opening of a gap by changing
charge carrier density from the metallic regime ($n =
3.5\times10^{11}~\text{cm}^{-2}$) to the CNP. At a temperature of
1~K we measure a resistance increase from 5~k$\Omega$ up to
14~M$\Omega$. The observation indicates the gapped ground state of
the studied bilayer graphene with a value of 6.8~meV. Measurements
in tilted magnetic field showed that the resistance at the CNP
decreases with an increasing of magnetic field. Based on this we propose a possible
scenario of the symmetry breaking in this bilayer graphene sample:
Spontaneous valley splitting at zero magnetic field followed by the
splitting of the spins originating from different valleys with
increasing of $B$. Both, the gap value and its weak
linear decrease with $B$, supports LAF as the ground state of the
studied sample.

\section{Experimental details}

Suspended bilayer graphene devices were prepared using an acid free
technique.\cite{Tombros1, Tombros2} We deposited highly ordered
pyrolytic graphite on $n^{++}$Si/SiO$_2$ wafer (500~nm thick) which
is covered with an organic resist LOR (1.15~$\mu$m). A standard
lithography procedure is performed in order to contact bilayer
graphene flakes (determined by their contrast in optical microscope)
with 80~nm of Ti/Au contacts. A second electron beam lithography
step is used to expose trenches over which graphene membrane becomes
suspended. To achieve high quality devices we use current annealing
technique by sending a DC current through the membrane (up to
1.1~mA) at the temperature of 4.2~K. While ramping up the DC
current, simultaneously, we keep track of the sample resistance.
Once the resistance reaches values in the order of 10~k$\Omega$s we
stop annealing and check the gate voltage dependence. We repeat this
procedure till the appearance of a sharp resistance maximum at the
CNP located close to zero $V_g$. More details on the current
annealing procedure can be found in Tombros $et~al.$\cite{Tombros2}
The studied device was 2~$\mu$m long and 2.3~$\mu$m wide. All
measurements were performed in four-probe geometry
(with contacts across the full width of graphene) at
the temperatures from 4.2~K down to 300~mK. The four-probe method
allows to eliminate contact resistances. As discussed
below the resistance measurements consist of a superposition of
longitudinal magnetoresistance ($\rho_{xx}$) and Hall-resistance
($\rho_{xy}$). The carrier density in graphene is varied by
applying a DC voltage ($V_g$) between the back gate electrode
($n^{++}Si$) and the graphene flake. Based on serial capacitors
model a unit capacitance of the system is 7.2~aF$\mu\text{m}^{-2}$,
which relates gate voltage with density as $n = \alpha V_g$ where
$\alpha$ is a leverage factor of $\alpha =
0.5\times10^{10}\text{cm}^{-2}\text{V}^{-1}$. The typical current we
use is around 1~nA. See Appendix.

\section{Temperature dependence and Quantum transport}

\begin{figure}
\includegraphics[width=8.6cm]{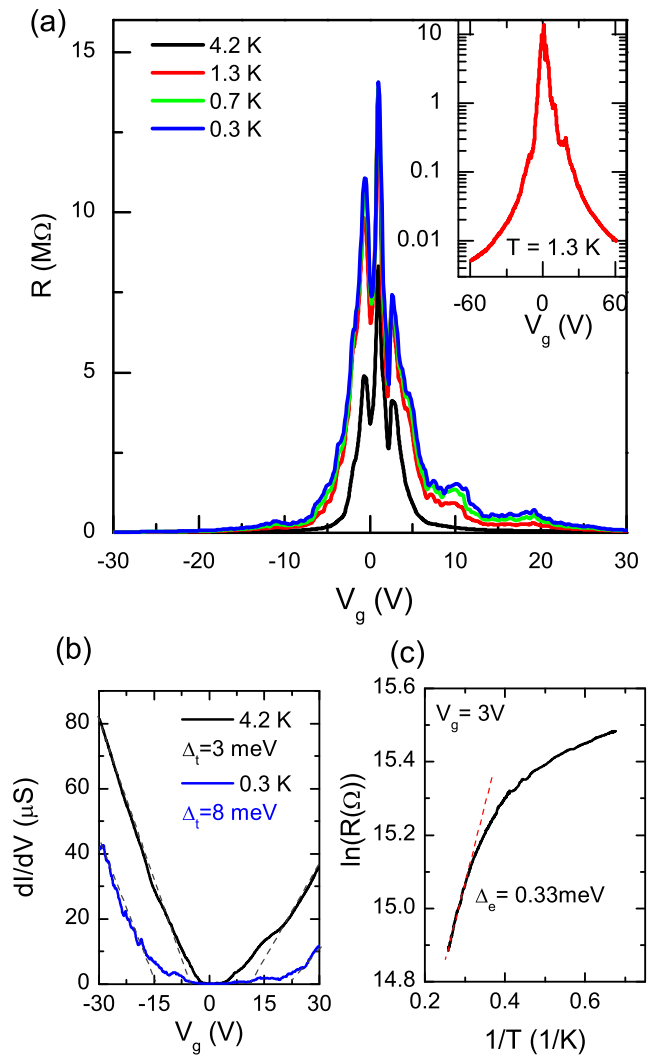}
\caption{\label{fig:Fig1}(Color online) 4-Probe resistance of the suspended bilayer graphene.
a) Gate dependence of the sample at the temperatures of 4.2~K (in black), 1.3~K (in red), 0.7~K (in green),
0.3~K (in blue). Inset: Resistance at 4.2~K in log-scale showing the dramatic change from the CNP to metallic regime;
b) Transport gap extraction at 4.2 and  0.3~K. The energy gap in bias direction
is highlighted by the conductance crossover (fitted with dashed lines) at zero. The values of the transport gap are 3~meV (4.2~K)
and 8~meV (0.3~K); c) An Arrhenius plot of the resistance. The value of an extracted thermal gap is 0.33~meV
}
\end{figure}

Our pristine samples are strongly p-doped with the CNP situated
beyond 60~V and a metallic resistance of a few hundreds of $\Omega$s
over the entire voltage range. Therefore we perform current
annealing technique in order to obtain high quality devices. In
contrast to previous samples, in which each next step of current
annealing tend to cause sharper change in the resistance values
within the scanned region of $V_g$, the discussed bilayer sample
already shows after the first current annealing step a high
resistive region around the CNP (not shown). The next annealing step
(1.1~mA) moves the charge neutrality point down to $V_g$~=~3~V.
However, surprisingly the resistance around CNP becomes 14~M$\Omega$
and reduces down to 5~k$\Omega$ in the metallic regime at
$V_g~=~\text{-60~V}$ (Fig.~\ref{fig:Fig1}a, Inset). This fact points
toward opening of a gap.
 The temperature dependence of the membrane from 4.2~K to 300~mK is shown in Fig.~\ref{fig:Fig1}a.
There is an essential change of about 6~M$\Omega$ in the maximum
resistance ($R_{max}$) from 4.2~K down to 1.3~K, however further
lowering of temperature does not change $R_{max}$ much. From an
Arrhenius plot of the resistance at CNP (Fig.~\ref{fig:Fig1}c) we
can extract a thermal excitation gap of 0.33~meV.\cite{Xia} The
saturation of resistance at lower T can be explained by a variable
range hopping with different temperature dependence. We would like
to point out that our excitation current value of 1~nA gave a
voltage drop of $\propto10$~mV at the CNP, which is much higher than
$kT$ energy at measured temperatures (0.3~meV). Therefore one has to
be careful in comparing transport and thermal excitation gaps.


There might be a couple of scenarios for the observed gap formation
in the gate voltage dependence: (i)~A lateral confinement in
membrane, where energy levels are
\begin{equation}
E_{n}= \frac{\hbar^{2}k^{2}}{2m}\ = \frac{\hbar^{2}\pi^{2}}{2m W^{2}}{l^{2}}\,
\end{equation}
$W$~=~2.3~$\mu$m - width of the flake, $l$ is integer value.
However, first two levels have energies of $E_1$~=~1.3~$\mu$eV and
$E_2$~=~5.3~$\mu$eV, which is much lower than $k_BT$ at measured
temperatures. (ii)~True gap formation with zero density of states
within the gap and available states at the conduction and valence
bands. (iii)~Transport gap, accompanied by the observation of the
reproducible conductance oscillations in the region of suppressed
conductance. In such regime transport is limited by the quantum
confinement effect along the width (mainly originating from the
impurities).\cite{Molitor} (iv)~More complicated case, when the gap
value depends on the charge carrier density, $i.e.$ the energy of
the levels changes while being filled with carriers. This situation
might happen when the gap is induced by charge redistribution in
between layers, which would be influenced by the applied back gate
voltage. At the moment, we can not determine the exact gap type,
therefore, further analysis is performed assuming a transport gap
scenario, but keeping in mind that this gap value can depend on the
density.

In an analogy to graphene nanoribbon studies\cite{Molitor, Stampfer}
we extract the transport gap from the gate dependence of the sample
conductance as shown in Fig.~\ref{fig:Fig1}b). From a linear
approximation of conductance one gets a region of $\Delta V_g$,
where sample shows insulating behavior. This region
$\Delta V_g$ relates to the wave vector as $\Delta k = \sqrt{\pi
\Delta n} = \sqrt{\pi \alpha \Delta V_g}$. Taking into account the
quadratic dispersion of bilayer graphene, the corresponding energy
scale can be calculated as
\begin{equation}
\Delta E_{F}= \frac{\hbar^{2}k^{2}}{2m}\ = \frac{\hbar^{2}}{2m}{\pi \alpha \Delta V_{g}}\,
\end{equation}

From conductance graphs at different $T$ we find $\Delta E_{F} =
3$~$meV$ at 4.2~$K$ and $\Delta E_{F} = 8$~$meV$ at 0.3~$K$. The
values of the transport gap are comparable to the energy gap
(extracted in bias direction) values of single layer graphene
nanoribbons of 50-85~nm  wide,\cite{Molitor, Stampfer} where in contrast to our case the gap is created by lateral
confinement. The resistance value of 5~k$\Omega$ in metallic
regime, similar to regular graphene devices, serves as an additional
justification of excluding a lateral confinement as a cause of the
observed transport gap. We can calculate the mobility of the charge
carriers using a standard formula $\mu = 1/(eR_{sq}n)$, where
$R_{sq}$ is a square resistance of the sample and $e$ is elementary
charge. The mobility value
$\mu\propto~20,000~\text{cm}^2\text{V}^{-1}\text{s}^{-1}$ at $n =
3.5\times10^{11}~\text{cm}^{-2}$ corresponds to the value of high
quality bilayer graphene devices. Due to the symmetry of resistance
change around CNP (Fig.~\ref{fig:Fig1}b) and the fact that CNP
itself is situated around zero gate voltage
($V_g$~=~1.2~V), that corresponds to the density of $n
= 0.77\times10^{10}~\text{cm}^{-2}$ at 0~V, we can also exclude the
low quality "p-doped" regions close by the contacts (which can form
after current annealing) as the cause of the reported gap. In the
meantime, we can not exclude a charge inhomogeneity in the sample
bulk which might lead to the observed order of magnitude deference
between electrical and transport gaps, in analogy to nanoribbon
case.

\begin{figure}
\includegraphics[width=8.6cm]{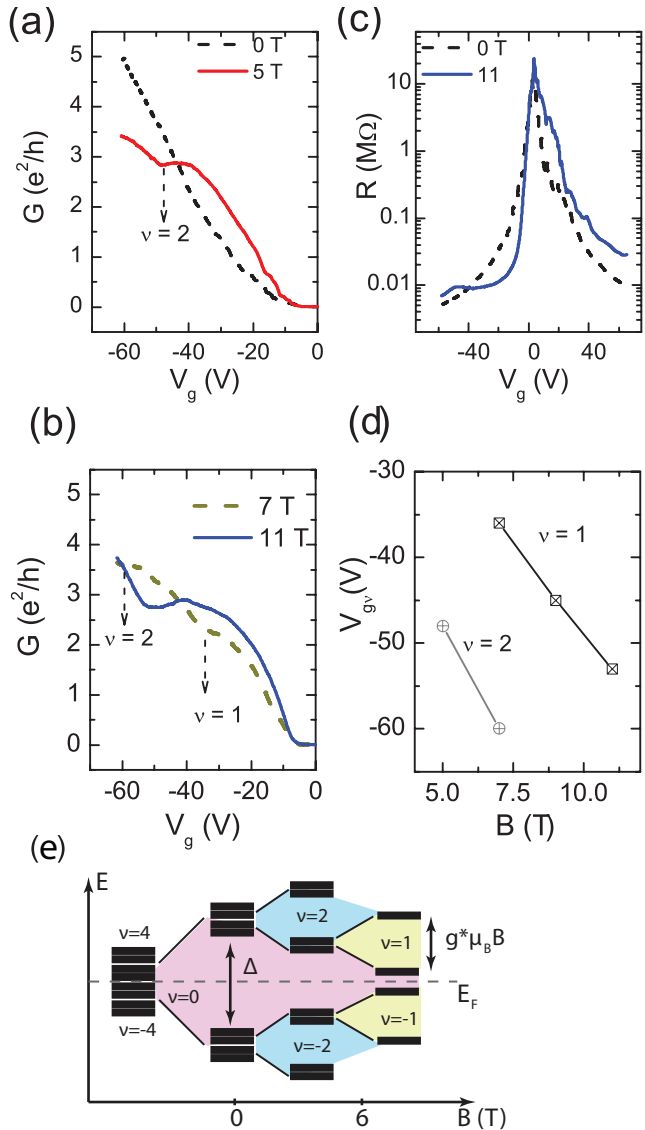}
\caption{\label{fig:Fig2}(Color online) Quantum transport at 1.3~K. a) Quantum Hall conductance of the suspended bilayer at zero and $B$~=~5~T;
b) Quantum Hall conductance of the suspended bilayer at $B$~=~7 and 11~T. The exact filling factors $\nu$ corresponding to the observed plateaus
are shown; c) Resistance of the sample in quantum Hall regime. d) Scaling of the filling factors position in gate voltage ($V_{g\nu}$) with magnetic field; e) LL hierarchy
of the symmetry-breaking of the lowest LL in bilayer graphene. Suggested scenario of the spontaneous valley splitting followed by the spin splitting at high $B$.
}
\end{figure}

Given the fact that the resistance values reach M$\Omega$s at the
CNP, it is already hard to establish quantum Hall plateaus in our
suspended bilayer device. However, we have achieved to observe
quantum Hall transport shown in Fig.~\ref{fig:Fig2}a,b). First
quantum Hall plateau appears at 5~T on electron side (red curve),
which we attribute to the filling factor $\nu = 2$. This plateau is
followed by the appearance of $\nu = 1$ at 7~T
(Fig.~\ref{fig:Fig2}b). The conductance values of the observed
plateaus deviate from the expected ones of $2e^2/h$ and $1e^2/h$, since they are affected by charge inhomogeneity. Therefore, we determine the exact values of the corresponding plateaus by
the scaling of their positions in density ($V_{g\nu}$) with magnetic
field $B$ (Fig.~\ref{fig:Fig2}d). As expected from $\nu = n/(eB/h)$
the scaling is linear with the leverage factor of $\alpha =
0.64\times10^{10}\text{cm}^{-2}\text{V}^{-1}$ for $\nu = 2$ and $1$.
In order to use the same $\alpha$ for both filling
factor sets (see Fig.~\ref{fig:Fig2}d) the slopes of $V_{g\nu}$
versus $B$; and $\nu$ values respectively, have to be twice as
different. Therefore, we have to point out that the linear scaling
will hold as well for a leverage factor of
$1.1\times10^{10}\text{cm}^{-2}\text{V}^{-1}$ in case we assume $\nu
= 4$ and $2$ as an observed sequence of plateaus. From previous
studies\cite{Ivan} we know that capacitance probed by the QHE in
graphene devices (especially in suspended samples) can be higher
than the geometrical value, due to the deviation from the plane
capacitor model. However, we attribute the observed plateaus to the
filling factors~2 and~1. As we noticed before,\cite{Tombros1,
Elferen} most of the time the current annealing procedure leads to
the formation of high quality annealed regions connected in series
with low mobility p-doped regions close to the contacts. Therefore
higher values of the conductance plateaus can be explained by a
"p-doped" slope, which increases with magnetic field $B$. This can
be also the reason of the absence of resistance quantization in the
electron-side (Fig.~\ref{fig:Fig2}c). Assuming $\mu B
\gg $1 for the formation of QHE plateaus,\cite{Bolotin1} our
observation implies a lower bound for the mobility of
2,000~$\text{cm}^2\text{V}^{-1}\text{s}^{-1}$.

To summarize our QH transport results: At this point we have shown
that a zero-field gap opens at the CNP in the studied graphene
bilayer. This observation points out on a possible symmetry breaking
of the ground state in bilayer graphene. The application of $B$ does
not restore the broken symmetry and brings the systems in to the QH
regime. In Fig.~\ref{fig:Fig2}e) we show the hierarchy of the
splitting of the 8-fold degenerate lowest Landau level in applied
$B$.\cite{Zhao1} The development of the level structure with $B$
will be specified and discussed in section 4. In the meanwhile, if
we assume that at $B$~=~0~T one of the degeneracies is already
lifted, then, with increasing field, one can expect quantization at
the filling factors $\nu = 0$ and $4$ followed by $\nu = 2$ and $1$.
However, if the initial symmetry breaking is strong enough and the
scanned window in energy is limited ($V_g$), then one can expect
quantization at $\nu = 2$ followed by $\nu = 1$. This described
hierarchy of levels splitting and sequence of plateaus will be
observed independent on either valleys or spin splitting first.

\section{Resistance at the CNP in tilted magnetic field}

\begin{figure}
\includegraphics{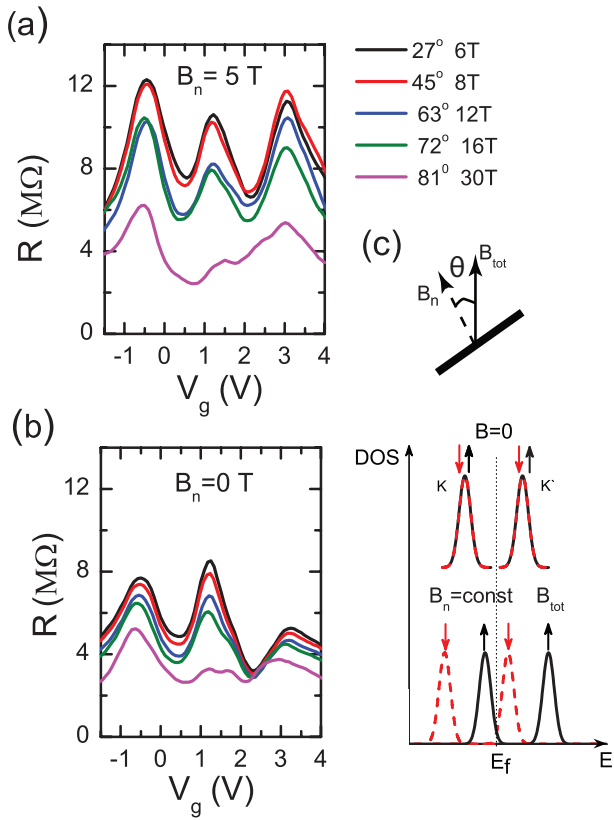}
\caption{\label{fig:Fig3}(Color online) a) Behavior of the resistance at the charge neutrality point at fixed $B_n$ and increasing $B_{tot}$. From top to bottom angle and total field: 27$^{o}$ (6~T), 45$^{o}$ (8~T), 63$^{o}$ (12~T), 27$^{o}$ (6~T), 72$^{o}$ (16~T), 81$^{o}$ (30~T)
b)Behavior of the resistance at the charge neutrality point when $B$ has only in plane field component($\theta$ = $90^{o}$).
c) Suggested scheme of the spontaneously split valley followed by spin splitting induced by $B$.
}
\end{figure}

In order to clarify the nature of the gapped ground state of bilayer
graphene and its evolution in magnetic field we perform a tilted
magnetic field experiment. In tilted experiments the total magnetic
field ($B_{tot}$) can be separated from its normal (to the plane of
the sample) component: $B_n =~ B_{tot}cos\theta$, where $\theta$ is
an angle between these two vectors (Fig.~\ref{fig:Fig3}c). This
procedure allows us to distinguish between the orbital effect (QHE)
and pure Zeeman energy, which has to scale with $B_{tot}$
value.\cite{Zhao1,Zhang3,Kurganova1}

All measurements presented below were performed at a temperature of
1.3~K. The application of the magnetic field perpendicular to the
sample plane leads to an increase in the resistance at the CNP, as
it is expected for a QH transport in the case of broken symmetry
states. To distinguish between normal component and total $B$ we
perform a series of experiments with keeping $B_n$ fixed and
gradually increasing $B_{tot}$. As an example, in
Fig.~\ref{fig:Fig3}a) we show a change in $R_{max}$ at $B_n$~=~5~T
and $B_{tot}$ increasing from 6~T up to 30~T for different angles
$\theta$. The actual maximum of the resistance consists
of three peaks: highly resistive in the middle ($V_g$~=~1.2~V) and
two side peaks at the gate voltage at -0.5 and 3~V. The total
magnetic field causes decrease in the resistance and the middle peak
starts splitting into two peaks (or developing minimum in resistance
at the CNP) when $B_{tot} > $~6~T for studied values of $B_n$. We
observe exactly the same behavior in the experiment when $B_n$~=~0
and applied field is in parallel to the graphene membrane: maximum
of the resistance goes down and develops a local minimum at the CNP
(Fig.~\ref{fig:Fig3}b). We attribute this change with an increase of
the total magnetic field. That fact that resistance changes with
$B_{tot}$ indicates that observed effect is not a simple quantum
localization due to inhomogeneity in the sample.

\begin{figure}
\includegraphics[width=8.6cm]{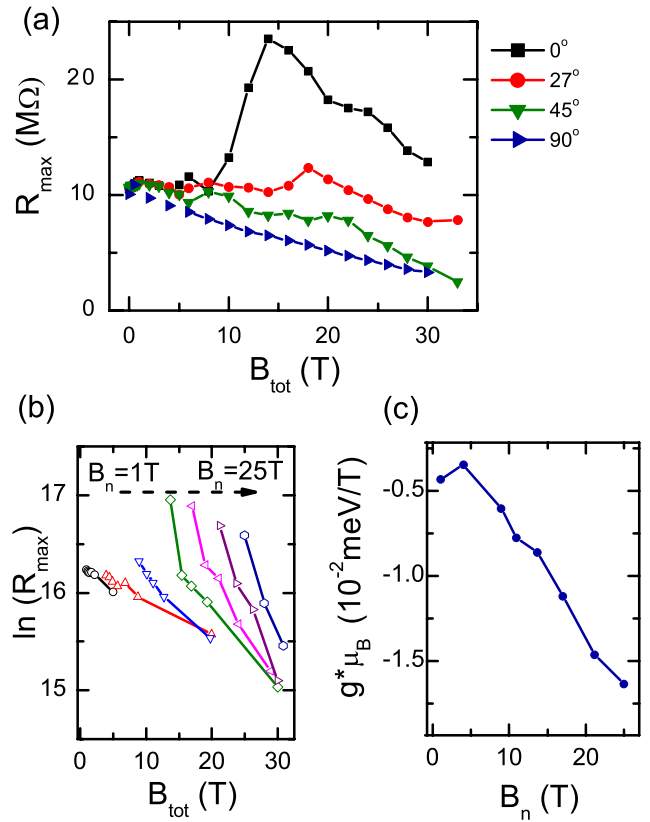}
\caption{\label{fig:Fig4}(Color online) (a) Change in the $R_{max}$ of the middle peak with total magnetic field $B_{tot}$.
 (b) ln($R_{max}$) as a function of $B_{tot}$ at different $B_n$. The values of $B_n$ from left to right 1, 4, 9, 13.7, 17, 21.2, 25~T.
 (c) The slope of the linear fit from Fig.4b) as a function of normal component $B_n$.}
\end{figure}

All three maxima around the CNP decrease in their
resistance in applied parallel $B$. However, only the middle maximum
at $V_g$~=~1.2~V shows clear scaling with the total magnetic field
($B_{tot}$) at different tilted angles $\theta$
(Fig.~\ref{fig:Fig4}a). As one can see in the case of $B_{tot} =
B_{n}$ ($\theta = 0 $, black curve in Fig.~\ref{fig:Fig4}a) the
resistance keeps on increasing up to around 14~T; further increase
in magnetic field brings $R_{max}$ to lower values
(Fig.~\ref{fig:Fig4}a). Once the non zero angle is introduced the
common trend for $R_{max}$ is a decrease.

We suggest that the behavior of the middle peak is
caused by the many-body effect and can be explained by the Zeeman
splitting closing the spontaneous gap. The hierarchy of energy
levels is depicted in Fig.~\ref{fig:Fig2}e). Once $B$ value is high
enough the LLL is split in to 4 levels, each two-fold degenerate. If
we assume that the latter degeneracy is spin, then after the
appearance of plateau associated with filling factor $\nu = 1$ we
expect the value of the ground state gap $\Delta$ to be lowered by
spin splitting coupled to $B_{tot}$. Here we would
like to emphasize, that we do observe appearance of $\nu = 1$ and
minimum of resistance at the CNP in similar magnetic field $B_{tot}
> 7$~T. In a simplified way we describe resistance value at the CNP point as
\begin{equation}
\ln{R_{max}} \propto \Delta/(kT)-g^{*}\mu_{B}B_{tot}/(kT),
\end{equation}
where $g^{*}$ is an effective $g$-factor including exchange electron
interaction  and a Landau level
broadening.\cite{Nicholas,Kurganova1,Elferen} The change in ln($R$)
versus $B_{tot}$ at fixed $B_n$ values is shown in
Fig.~\ref{fig:Fig4}b). This dependence can be the best described as
linear. The slope and y-intercept of the linear fit of
Fig.~\ref{fig:Fig4}b) give the values of $\Delta$ and $g^{*}\mu_B$.
Surprisingly, these both contributions scale with $B_n$ component.
In Fig.~\ref{fig:Fig4}c) we show $g^{*}\mu_B$ values versus $B_n$.
Despite the fact that the scaling seems like linear, plotting the
slope as a function of $\sqrt{B_n}$ does seem like fitting as well
(not shown). $\Delta$ value increases with $B_n$ from 1.4~meV at
$B_n$~=~1~T up to 1.7~meV at $B_n$~=~25~T (not shown). This $\Delta$
is of the same order as the measured transport gap
(which can overestimate a real energy gap) and also
corresponds to the theoretically predicted gap of 1.5-30~meV for the
excitonic instability.\cite{Nandkishore2, Jung1, Zhang3}

In summary, tilted magnetic field experiments show that the
resistance at the CNP of studied gapped bilayer graphene decreases
linearly with the total magnetic field component. This points to a
many-body effect and weak reduction of the gap in
applied magnetic field. The developed minimum in the resistivity in
Fig.~\ref{fig:Fig3} can be explained by the overlapping of spin-up
and spin-down levels from the adjacent Landau levels due to Zeeman
splitting in applied $B$.\cite{Nicholas} However, from
our experiments the estimated $g^{*}<$~0.2, which is very low for
spin splitting. In addition, although the resistance decreases in
parallel field, the $R_{max}$ value does not change an order of
magnitude. This behavior in $B$ is consistent with the layer
antiferromagnetic state as a ground state of studied bilayer
sample.\cite{Zhang3} Since in this state the top and bottom layers
host spins with opposite orientations, their interaction with
applied $B$ can not be described as a simple Zeeman splitting. Next
to it, our results also open an additional question: What is the
role of exchange energy and level broadening $\Gamma$ in LAF state?
Naively, scaling of $g^{*}\mu_B$ with $B_n$ can be understood from
their dependence on level broadening $\Gamma$. The $\Gamma$ value
scales with $\sqrt{B_n}$, meaning the bigger $B_n$ the smaller
$B_{tot}$ is needed to observe level's overlapping. In reality the
situation can be much more complicated including possible exchange
mechanisms we do not understand yet. This is also supported by the
fact that the ground state gap $\Delta$ depends on $B_n$ as well.

Based on these results we suggest a possible scenario of symmetry
breaking in high quality bilayer graphene (Fig.~\ref{fig:Fig2}e and
Fig.~\ref{fig:Fig3}c). First splitting is caused by valleys, which
results in the observed transport gap. An application of magnetic
field induces spin splitting of both $K$ and $K'$ levels. When $B$
is high enough then the energy of spin-up level from $K$ will start
approaching the spin-down level from $K'$. The overlapping of the
levels will cause a decrease in the resistance at the charge
neutrality point. Since we do observe transport gap in our sample,
we exclude nematic phase transition. In addition to
this, the response of the sample in tilted $B$ fits to the LAF
state. The cause of the valley splitting can be a
combination of two effects: electron-electron interaction (which
determines the $B$ field behavior of the middle resistance maximum )
and a contamination of the sample surface with charged impurities
which break inversion symmetry (via introduction of electrical
field).\cite{Castro}









%




%











%






\section{Conclusions}

We report a transport gap of 3~meV in suspended bilayer graphene at
4.2~K, which increases with decreasing of temperature.  The sequence
of appearance of the QHE plateaus at the filling factor $\nu = 2$
followed by $\nu = 1$ supports a suggestion that the observed gap
caused by the symmetry breaking. Measurements in the tilted magnetic
field indicates that the resistance at the CNP
shows weak linear decrease with the total magnetic
field component. Based on this we propose a possible scenario of the
symmetry breaking in the investigated bilayer graphene: Spontaneous
valley splitting at zero magnetic filed followed by the splitting of
the spins originating from different valleys with increasing of $B$.
The gap value and weak response of the sample to applied magnetic
field corresponds to the predicted spin polarized layer
antiferromagnetic state as a ground state of the investigated
sample. The observed non-trivial dependence of the gap value from
the normal component of $B$ suggests possible exchange mechanisms in
the system.

\begin{acknowledgments}
We would like to thank B.~Wolfs, M.~de~Roosz and J.G.~Holstein for
technical assistance. We also thank M.H.D.~Guimar\~{a}es for useful
discussions; and I.J.~Vera-Marun for creating a soft-wear program
for current annealing. This work is supported by NWO (via TopTalent
grant), FOM, NanoNed and the Zernike Institute for Advanced
Materials.
\end{acknowledgments}

\section{Appendix}
\begin{figure}
\includegraphics[width=8.6cm]{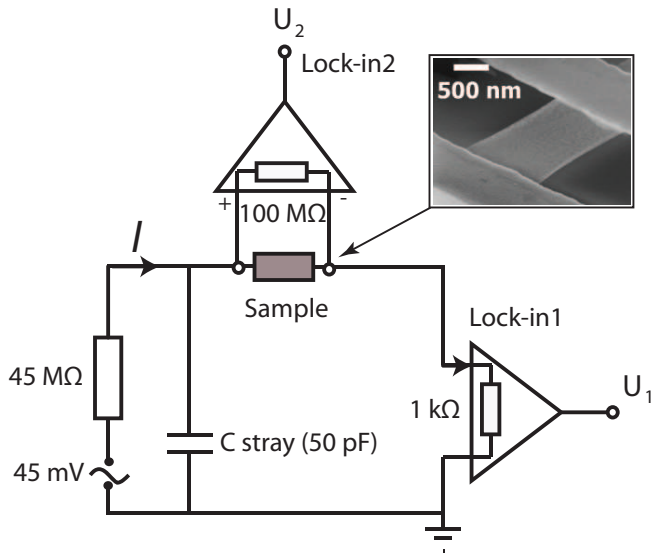}
\caption{\label{fig:Fig5}(Color online) Electrical scheme of the setup we use to perform our measurements.
Inset: Scanning electron micrograph of a typical suspended bilayer membrane in between two contacts.
}
\end{figure}

In order to minimize self-heating in graphene at the high resistive
CNP we used the following scheme (Fig.~\ref{fig:Fig5}). An AC source
maintained a fixed voltage amplitude of 45~mV (1.87~Hz frequency)
across the sample in series with 45~M$\Omega$ resistor. The current
through the sample is monitored by the lock-in1, whose output $U_1$
is proportional to the current flowing in the circuit ($U_1 = I
\times 1 \text{k}\Omega$). Simultaneously, the four probe voltage
across the sample ($U_2$) is phase detected by another lock-in2
connected through the preamplifier having an input resistance up to
100~M$\Omega$. Then the resistance of the sample is determined by $R
= 1\text{k}\Omega  \times U_2/U_1$. The power dissipating in the
sample is $P = U^2_2/R$. Therefore, assuming that maximum $V_2$ is
already reached ($\propto10$~mV), with increasing of $R$ the
dissipation in the sample will be decreasing.

\end{document}